\documentclass[aps,prd,showpacs,floatfix,preprint]{revtex4}
\usepackage{amsmath,bm}
\usepackage{graphicx}
\usepackage{epsfig}
\begin{document}

\title{Probing the metastability of protoneutron star with hyperon in core-collapse supernova} 
\author{Sarmistha Banik}
\affiliation{BITS Pilani, Hyderabad Campus, Hyderabad 500078, India}

\begin{abstract}
We investigate
the role of hyperons in the dynamical collapse of a non-rotating
massive star to a black hole(BH) using one dimensional general relativistic $GR1D$
code. We follow the dynamical formation and evolution 
of a protoneutron star (PNS) to a black hole 
using various progenitor models, adopting a hyperonic equation of state 
(EoS) generated by Shen et. al. 
We compare the results with those of nuclear EoS by Shen et. al. and
understand the role of strange hyperons in the core collapse supernova. 
We discuss the neutrino signals that may be used as a probe to core collapse.
Further, an exotic EoS can support a much lower maximum mass cold neutron 
star compared to PNS. In this regard, we also
study the metastability of PNS in the presence of hyperon in the long-time 
evolution of the progenitors, relevant to supernova SN1987A.
\pacs{26.60.Kp, 26.50.+x, 26.60.-c, 14.20.Jn}
\end{abstract}
\maketitle

\section{Introduction}

The study of hot and dense matter relevant to neutrons stars is an interesting problem. 
Apart from conventional nuclear matter, the neutron star core might contain 
exotic matter such as hyperons, quarks and Bose-Einstein condensates of 
antikaons at higher densities \cite{Gle}. It is obvious that the inclusion of 
strange degrees of freedom softens the equation of state (EoS). A stiffer EoS 
can result in higher maximum mass neutron stars. A soft EoS, on the other hand,
favors lower maximum masses compared to the stars having nucleonic degrees of freedom only. The recent measurement of the Shapiro delay in the radio
pulsar PSR J1614-2230 which yielded a mass of $1.97\pm0.04M_{\odot}$\cite{demo} 
and the mass measurement of PSR J0348+0432 having mass $2.01\pm0.04 M_{\odot}$ 
\cite{anto}, put important constraint on the neutron star mass and may rule out 
most of the soft EoS . However, it is at present not possible to rule out any 
exotica with this observation as many model calculations including $\Lambda$ 
hyperons and/or quark matter could still 
be compatible with the observations \cite{shenY,Lat10,mass}. Many of these approaches are parameter dependent, for example the EoS with $\Lambda$ hyperons 
\cite{oertel} are compatible with the benchmark of $2{M_\odot}$. 
 

EoS provides crucial nuclear physics input to the core collapse 
supernova simulations. To generate a finite temperature EoS table which covers 
a wide range of density ($10^4 -10^{15} g/cc$), temperature ($0-100$MeV) and
composition (proton fraction $0-0.6$) is indeed a challenging task. Also 
sub-saturation density region containing nuclei and non-uniform nuclear matter 
relevant to crust of neutron stars should be carefully matched with the 
high-density EoS. Mainly two sets of EoS $-$ Lattimer-Swesty (LS) \cite{Lat} 
and Shen, Toki, Oyamatsu and Sumiyoshi (Shen) EoS\cite{shen} are widely used for the supernovae simulations. These contain non-strange particles like neutrons, protons, alpha-particles and nuclei. LS EoS is based on the
non-relativistic liquid drop model while the Shen EoS is computed in the framework of relativistic 
mean-field (RMF) with TM1 parameter set \cite{Sug} taking into consideration the Thomas-Fermi approximation. Both the models are worked out in single representative nucleus and $\alpha$-particles of light clusters; no shell effect is considered. Other notable nucleonic EoS are based on
thermodynamically consistent nuclear statistical equilibrium and RMF models \cite{hempel, furusawa}, virial expansion methods \cite{horeq}. The first EoS with non-nucleonic degrees of freedom was presented by Ishizuka et. al. \cite{ishi08}. They studied the emergence of the full baryon octet in the dynamical collapse of a massive static star to a black hole(BH) \cite{ishi}.  This EoS was recently utilised to study the behaviour of black hole formation and neutrino emission with hyperons and/or pions in \cite{naka11}. Recently
stellar core collapse simulations were reported with additional pions and $\Lambda$ particles in LS EoS \cite{oertel}. Shen et. al. extended
their nuclear EoS \cite{shen} to include hyperon degrees of freedom in the RMF framework \cite{shenY}, which we eventually use for our supernova simulations and label as np$\Lambda$ EoS.

Neutrinos are very important observables for the core collapse supernova explosions.
The massive stars at their final journey implode so quickly that the inner 
core rebounds; a shock wave is said to form \cite{bethe}. After the core bounce,
trapped neutrinos diffuse out of the core and escape the surface of the star, 
which can be recorded by the detectors on earth. The neutrinos carry off most 
of the energy; the shock soon loses its power and stalls after traversing a 
few hundred kilometers \cite{bethe}. The neutrino signals cease. There are 
different ideas to revive the shock, which could trigger a delayed supernovae 
explosion.  One such idea is a quark-hadron  phase transition in the high 
density core of the compact star \cite{sagert11}. This would be manifested by a 
subsequent set of neutrino burst.

SN1987A, since its discovery, has become the most studied star remnant in history and has provided great insights into supernovae and their remnants.
Observation of a burst of neutrino signal for at least 12s after the 
explosion strongly supports to the scenario that a proto neutron star (PNS) 
was initially present in the core which cooled via neutrino emission and 
collapsed later.
The fading light curve also lends support to this picture. During deleptonisation, the hot and neutrino-trapped PNS may end up either to form a cold neutron star (CNS) or a BH.  Till date astronomers have 
not been able to find a neutron star in the remnant of this type II supernova. 
However, the existence 
of a metastable PNS at least 12s after the explosion indicates to some deep 
physical implications \cite{meta, brown}. 

It is believed that the fate of the compact object depends on the EoS and the amount of infalling material. The general relativistic calculations limit a maximum mass that can be sustained by an EoS \cite{ruff}. There are two possible scenarios of formation of BHs after the supernova explosion. If the BH formation time is comparable to accretion time, the PNS must have accreted sufficient mass on that time scale to overshoot the maximum limit set by the EoS \cite{brown2}. 
However, a delayed BH formation indicates the metastability in neutron stars on the deleptonisation time-scale which is consistent with the observation of SN1987A. Strangeness could be the reason for such delay in BH formation \cite{prak_meta}. Strange matter, that is believed to exist in the high-density core of neutron stars \cite{Gle} might already exist in the early post-bounce phase of a core-collapse supernova.  We explore if strange
$\Lambda$ hyperons can drive the BH formation in the long time evolution of the PNS i.e. 12 seconds as was in SN1987A.

It has been reported by several authors that the  PNS made up of nucleons 
have a slightly smaller maximum mass than the neutrons stars 
\cite{meta,third, schaf}.
Bethe and Brown calculated the maximum mass of the CNS which is known as the Bethe-Brown limit ($1.56 M_\odot$)\cite{brown}. PNS with exotic matter have maximum mass larger than CNS and Bethe-Brown limit, a reversal of the conventional nuclear matter scenario \cite{meta, brown}.  Since baryonic mass is conserved, once sufficient thermal support is lost with cooling, the PNS becomes metastable
and it would collapse to a low mass BH. The delayed collapse to BH was studied using a hyperon EoS earlier \cite{Keil, Baum}. However, the hyperon
EoS used in these cases were not a state of art EoS like Shen hyperon EoS \cite{shenY}. 

In this paper, we report the influence of hyperons on the BH formation using the spherically-symmetric
general relativistic hydrodynamic code, $GR1D$\cite{Ott}, designed to follow the evolution of stars beginning from the
onset of core collapse. We adopt two sets of the Shen EoS- for nucleon (np) 
\cite{shen} and $\Lambda$ hyperon (np${\Lambda}$) degrees of freedom \cite{shenY}.  
We comment on the neutrino signal that might be observed as a result of phase transition from nucleonic to hyperonic matter.
Finally, we investigate the problem of metastability of the PNS in SN1987A and 
whether hyperons can delay the BH formation for a while ($\sim12$ s) 
or not. The paper is arranged as follows. In Section 2, we briefly describe the EoS and GR1D code. Section 3 is devoted to result discussion. 
Finally we summarise in Section 4.

\section{The EoS and the numerical simulation}

We use the nuclear and $\Lambda$ hyperon EoS by Shen et. al \cite{shen, shenY} for our simulations.  Shen nuclear EoS is based on a
relativistic mean field model at  intermediate and  high densities ($\rho > 10^{14.2}$ g/cc).
At low temperature (T$\le 14$MeV), and $\rho < 10^{14.2} g/cc$,
Thomas Fermi approximation is used. 
The non-uniform matter at low temperature and density is modeled to consist of
free nucleons, $\alpha$ particles and heavy nuclei whereas at extremely low density ($\rho < 10^{10}$ g/cc) and
finite temperature uniform nucleon gas of n, p, $\alpha$ particles  is
considered. Leptons are treated as uniform
non-interacting relativistic particles and their contributions are added separately.
Minimisation of free energy is done both for non-uniform matter and uniform nucleon gas at low density. For $\rho > 10^{14.2}$ g/cc,
the calculation has been done in finite temperature, field-theoretical RMF model in which the interactions
among baryons are mediated by the exchange of $\sigma$, $\omega$ and $\rho$ mesons. In this case, the calculations are done with the TM1 
parameter set \cite{Sug}, that are obtained by fitting the experimental data for binding energies and charge radii of heavy nuclei. 
With the TM1 parameter set, the nuclear matter saturation density is
$0.145 fm^{-3}$, the binding energy per nucleon is 16.3 MeV, the symmetry energy is 36.9 MeV and the 
compressibility is 281MeV \cite{Sug}.  
One advantage of the RMF model is that it can include hyperons systematically.
For their EoS including  $\Lambda$  hyperons, Shen et. al. use the experimental mass 
$M_{\Lambda}= 1115.7$ MeV \cite{shenY}. 
The coupling constant for $\Lambda$ hyperon-vector meson interaction is taken based on the naive quark-model and 
whereas the coupling constant for $\Lambda$ hyperon-scalar meson interaction is determined by fitting experimental 
binding-energies  data for single-$\Lambda$ hypernuclei \cite{shen06}.
$\Lambda$ hyperons appear when the threshold condition,
$\mu_{n}=\mu_{\Lambda}$ is satisfied at higher density, where $\mu_n$ and $\mu_{\Lambda}$
are the chemical potentials of neutron and $\Lambda$ hyperon  respectively.
Other hyperons, $\Xi$ \& $\Sigma$ are excluded due to their relatively higher threshold and lack of experimental data \cite{shenY}.

We use the open source code $GR1D$  \cite{Ott} for the supernova simulations. 
$GR1D$ is
a spherically-symmetric, general-relativistic Eulerian hydrodynamics code for
low and intermediate mass progenitors. It is designed to follow the evolution of stars beginning from the onset of core
collapse to BH formation and makes use of several microphysical EoS.
Neutrino effects are crucial in stellar collapse, they are the source of both 
cooling as well as heating \cite{Ott}. Here neutrinos
other than electron type ($\nu_e$) and anti-electron ($\bar\nu_e$) are grouped 
as  $\nu_x$. So, three sets of neutrino species, $\nu_e$, 
$\bar\nu_e$ and $\nu_x=\left(\nu_\mu, \bar\nu_\mu,\nu_\tau, \bar\nu_\tau\right)$ are considered. Neutrino emission takes place 
when electrons are captured by free or bound protons leading to fall of lepton number. In this code, lepton fraction is parameterized as a
function of density according to Liebend\"orfer's prescription
\cite{Lieb}.  However, post bounce, it can not capture the effect of
neutrino cooling, deleptonisation and neutrino heating. Hence a
3-flavor, energy-averaged  neutrino leakage scheme is adopted  \cite{Ott}. This captures
the effects of cooling.  The leakage scheme provides approximate energy and number
emission rates. Neutrino heating is included via a
parameterized charged-current heating scheme based on Ref. \cite{janka}. One can
adjust the neutrino heating in the simulation
through a scale factor $f_{heat}$ as discussed in Ref \cite{Ott}. We take 
$f_{heat}=1$, if not mentioned otherwise.

\section{Results \& Discussion}

We solve the Tolman-Oppenheimer-Volkov equation for zero temperature (T=0) 
EoS of neutron stars (NS) assuming neutrino-less 
$\beta$ equilibrium.  The maximum mass of the neutron star  for np EoS is 
2.18$M_{\odot}$, whereas for np${\Lambda}$ EoS, the maximum mass reduces to
1.82$M_{\odot}$. The corresponding radii are 12 and 12.5 km respectively.

We perform simulations with $f_{heat}=1$ for several progenitor models of 
Woosley et. al 
\cite{woo} using GR1D \cite{Ott}  for Shen EoS- np as well as and 
np${\Lambda}$ EoS \cite{shen, shenY} and tabulate bounce time, BH formation 
time, 
maximum baryonic and gravitational mass of PNS in Table 1. 
The black hole formation time is quoted in post-bounce, not in real timeline, 
unless mentioned otherwise.
We observe that the maximum mass is higher than that of NS. 
We run our simulations until post bounce 3.5s assuming they might have 
exploded in nature by this time \cite{Ott}.  When accretion pushes PNS over 
its maximum mass, a BH is formed.  
In Fig 1.  we plot  the temporal evolution of the baryonic and gravitational 
mass of PNS for progenitor models, $40M_{\odot}$ 
and $23M_{\odot}$. The bounce corresponds to the spikes at 
real timeline $t_{bounce}=0.273$s and $0.266$s respectively 
for the two progenitors, which we take as post-bounce t=0 in the figure. 
The value of $t_{bounce}$ is same for the np and 
np${\Lambda}$  EoS, as the contribution of hyperon is not important at that time as we 
would see later in Fig 6.  A BH formation is marked by a blow-up 
and a spike in the gravitational mass. The upper set of lines of Fig.1 is for 
the $40M_{\odot}$  progenitor. The negative gravitational energy accounts for 
the lower gravitational mass compared to 
the total mass of all its constituents, i.e. baryonic mass. 
We use dashed dark lines (full, colour online) for np${\Lambda}$(np) 
EoS. For the np EoS, the BH is formed with baryonic mass  2.706$M_{\odot}$ at 
1.085s
post-bounce, whereas for np${\Lambda}$ EoS (the dashed lines) this happens much earlier 
at 0.565s post-bounce for baryonic mass 2.384$M_{\odot}$. Strangeness 
degrees of freedom softens the EoS and in the process can support  less mass 
against gravity compared to non-strange stars. Continued
accretion thus overshoots the maximum mass early. Or in other words, the
stiffer EoS leads to larger post-bounce time to BH-formation.
We compare this with the results of a $23M_{\odot}$ progenitor (the lower set of lines in Fig. 1). The PNS mass 
is much less than that of $40M_{\odot}$ progenitor. Also, in this case the PNS 
accretes matter for a longer time until it blows as BH at 1.623(0.847)s post 
bounce for np(np${\Lambda}$) EoS, the maximum baryonic mass being 2.594(2.304)$M_{\odot}$. 
Here also ${\Lambda}$ hyperons is noted to hasten the BH formation.

\begin{table}

\caption{BH formation time and maximum mass of PNS(both baryonic and 
gravitational) for $f_{heat}=1$ for np and np$\Lambda$ EoS of Shen et al. 
\cite{shen,shenY} and different progenitor models of Wooseley \& Heger \cite{woo}}.
\begin{tabular}{|c|c|ccc|ccc|} 
\hline
&&\multicolumn{3}{c|} {np} &\multicolumn{3}{c|} {np${\Lambda}$}\\
Model&$t_{bounce}$& $t_{BH}$ &$M_{b, max}$&$M_{g, max}$& $t_{BH}$ &$M_{b, max}$&$M_{g, max}$\\
&(s)&(s)&$(M_{\odot})$&$(M_{\odot})$&(s)&$(M_{\odot})$&$(M_{\odot})$\\
\hline
S15WH07&0.174&2.810&2.192&2.039&2.814&2.183&2.031\\
S20WH07&0.232&2.360&2.420&2.243&1.612&2.212&2.083\\  
S23WH07&0.266&1.623&2.594&2.461&0.847&2.304&2.19\\   
S25WH07&0.235&2.060&2.481&2.299&1.377&2.225&2.104\\ 
S30WH07&0.203&2.809&2.306&2.142&2.260&2.20&2.06\\ 
S35WH07&0.241&2.121&2.620&2.420&2.125&2.591&2.40\\
S40WH07&0.273&1.085&2.706&2.554&0.565&2.384&2.336\\ 
S45WH07&0.262&2.104&2.612&2.448&1.171&2.661&2.483\\
S50WH07&0.190&2.510&2.285&2.129&2.114&2.19&2.057\\
S55WH07&0.172&2.460&2.279&2.120&2.260&2.162&2.030\\
S60WH07&0.188&3.060&2.158&2.011&3.212&2.142&1.999\\ 
S70WH07&0.221&2.860&2.339&2.167&2.197&2.162&2.030\\ 
S80WH07&0.210&3.060&2.556&2.094&2.113&2.134&2.001\\
\hline
\end{tabular}
\end{table}
Figs. 2 and 3 show the time evolution of central density ($\rho_c$) and 
temperature (T) respectively 
for the np and np${\Lambda}$ EoS in the two panels. Here we use dark 
lines(colour online)
to distinguish $40M_{\odot}$ and $23M_{\odot}$ progenitors.
The onset of BH formation is marked by a sharp rise in the value of $\rho_c$  
and T. The central density 
just before BH formation reaches about $10^{15}$ g/cc and temperature varies from 22 to 32 MeV for np to np${\Lambda}$ 
EoS for both the progenitors.  Owing to the hyperon emergence, the contraction of PNS is accelerated, which leads to
quicker rise in temperature and  central density. 

In Fig. 4, we compare the density profiles of PNS for np and np${\Lambda}$ cases in the two panels.
We use coloured lines (online) for $23M_{\odot}$ progenitor. For  $40M_{\odot}$ progenitor,
it is noted that the density rises from less than normal nuclear matter density 
($\rho_0\simeq 2.4 \times 10^{14} g/cc$) at the 
surface to a few times $\rho_0$ at the core.  The plateau in the
mid-radius region could be attributed  to strong thermal pressure
as evident in Fig. 5.  At core bounce, the central density is $1.4\rho_0$. 
With intense accretion, the central density shoots to $\sim 2\rho_0$ at 0.363s,
and $2.5\rho_0$ at 0.563s for the progenitor of $40M_{\odot}$ with np EoS 
[left panel]. 
The trend in the density profile remains similar for np and np${\Lambda}$. 
The central density is slightly above that of the np case at t=0.363s in the presence of $\Lambda$s, 
that just start appearing in the system. However, at t=0.563s, owing to a 
substantial amount  of $\Lambda$, the central density rises to almost $3.9\rho_0$, 
which is $\sim 2.8$ times its value at core bounce [right panel]. Next we discuss
the density profile for $23M_{\odot}$ progenitor that look similar to 
that of the $40M_{\odot}$ progenitor just after the bounce. However difference creeps up
with time and becomes much more pronounced in the presence of hyperons. For 
np(np${\Lambda}$) EoS, at 0.563s central density rises to $2.2(2.5)\rho_0$ 
compared to $2.5(3.9)\rho_0$ for $40M_{\odot}$ case. Just before BH 
formation (at $\sim 0.847$s) the central density reaches $3.9\rho_0$ for 
np${\Lambda}$ system. Central density remains $2.6\rho_0$ 
for np system at this instant until it reaches $17\rho_0$ when the BH is 
formed at 1.623s (not shown in graph).
 
We compare the evolution of temperature profiles in the absence/presence of $\Lambda$s in Fig. 5. 
In the both cases, temperature  attains a peak at the mid-radius region. 
For the $40M_{\odot}$ progenitor the peak rises from 66.8Mev at 0.363s to 
79.4Mev at 0.563s in np case [left panel]. This is due to accretion and
compression of shock heated material onto the PNS surface. At this region, the
thermal pressure support is enough to flatten the density profile.
In inner core ($\sim 6$km) the material is not shock heated, rather is
heated by adiabatic compression. The
temperature peak is further raised to 91.7MeV at 0.563s in the presence of 
$\Lambda$ hyperons[right panel]. When compared with a $23M_{\odot}$ progenitor,  the rise of temperature is not so quick. 
It increases upto 71.5(74.4)MeV at 0.563s for np(np${\Lambda}$) which accounts for the corresponding lower central density 
of Fig. 4.  But the temperature eventually reaches  80(169)MeV just before 
0.847s, when BH is formed for np${\Lambda}$ system. The temperature peaks to 167 
MeV before BH formation for np system at 1.623s(not shown in figure).

Next we compare the compositions of PNS in Figs. 6 and 7. In Fig. 6 we notice
that initially at core bounce the system consists of neutron and protons only;
hyperons appears first in the collapse at 0.16s after core bounce (assuming 
$10^{-3}$ considerable amount of fraction). The central density that was just 
above normal nuclear matter density (see Fig. 2) at bounce rises to 
$3.79\times10^{14} g/cc \simeq 1.6\rho_0$  at 0.16s after bounce. The 
temperature also increases to 16.26MeV. This is at par with our earlier 
findings that the appearance of $\Lambda$ hyperon is delayed until the matter 
density reaches at least $~2\rho_0$ at T=0 \cite{third, ddrh} and the threshold
density shifts to lower density with increasing temperature \cite{shenY}.
We display a snapshot of mass fraction for the $40M_{\odot}$ progenitor at
0.363s and 0.563s in the left panel of Fig. 7.
It is interesting to note that hyperons appear off-center owing to high
temperature, although density is still on the plateau. At 0.363s
after core bounce,  the abundance of $\Lambda$ becomes significant 
at R $\simeq 10$km,
as temperature is maximum there (Fig. 5 ). It even falls sharply after
reaching the peak due to fall in
temperature, only to rise at the core again owing to high density there.
At a later time, the high central density forbids it from dropping too low,
once it reaches the peak at mid-radius region.
Thus, $\Lambda$ becomes one of the  major components in the core.
Similar snapshots are drwan for $23M_{\odot}$ progenitor in the right panel 
at 0.363s and 0.840s.

We study the evolution of total neutrino-luminosity, that includes 
contribution from $\nu_e$, $\bar\nu_e$ and rest $\nu_x$ \cite{JPCS} and find a short 
neutrino burst ($\sim 1$s) before the PNS, born temporarily in a 
failed-supernova, terminates in a BH.
The resulting neutrino burst in np and np${\Lambda}$ cases are quite similar, 
differ only in  earlier termination of burst in the latter. 
The neutrino burst ceases at 1.085s and 1.623s post-bounce for np case for
$40$ and $23M_{\odot}$ progenitors respectively. 
The soft np${\Lambda}$ EoS lowers the critical mass
of PNS, thus accelerates the mass accretion onto it and triggers the
gravitational instability  at 0.565s and 0.847s post-bounce for $40$ and  
$23M_{\odot}$ progenitors respectively. 
However, no second neutrino burst is observed as was in quark-hadron phase
transition \cite{sagert11}. 
The quark EoS is stiff while the nucleon-hyperon phase transition being weak
first order or second order, yields a soft EoS.  
So, np${\Lambda}$ though triggers BH formation, fails to generate 
a second shock.

At this point we recall the observationobserved to be  of burst of 11 and 8 anti-neutrinos 
from SN1987A over the time period of $~12$s in the underground detectors of 
Kamiokande II in Japan and IMB in Ohio, respectively. All the evidence reveals 
that the supernova SN1987A lies at the position of a former, blue super-giant 
star, Sanduleak-69202  of mass around $20M_{\odot}$.
So, did SN1987A 
create a BH at the end of 12s, when the neutrino signals ceased or 
did it end up as a neutron star? In order to explain the observations, we 
follow the evolution of the progenitors of mass $\sim 20M_{\odot}$ for 12s. Since 
one-dimensional supernova models can not produce an explosion,  we increase 
the energy deposition by artificially raising the neutrino heating ($f_{heat}$).
The scaling factor $f_{heat}$ appears in the parameterized neutrino heating 
\cite{janka, Ott}. Until now, we used standard setting of $f_{heat}=1$ in our
calculations.  
Our motivation is to study if the shock can be revived through neutrino heating 
and whether metastability is a plausible mechanism with hyperons to collapse 
the PNS in a BH. Once sufficient matter 
is accreted onto the PNS, it might overshoot the maximum mass that can be supported by the EoS and eventually would collapse to a BH after a while.
Ott et.al found BH formation for 
some extreme models within 1s of the cooling phase \cite{ott_fail}. 
In Fig 8. we plot the shock radii for the $23M_{\odot}$ 
progenitors with different values of $f_{heat}$. For higher values of 
$f_{heat}$, the shock radii increase with time. We find $f_{heat}=1.37$ 
corresponds to the critical value required to cause a successful explosion. 
For $f_{heat}=1.35$, a BH is observed to be formed.
The PNS never ceases to accrete mass even for higher values of $f_{heat}$.
We do not notice any BH formation through metastability during 
20s after bounce.  We compare the density and temperature profiles with 
$f_{heat}>1$ to our previous simulations in Figs 9 and 10.
We do not notice any signature of blow-up, which could support the 
metastability and consequent BH formation. Even after 12 s, PNS 
remains stable. Thus at the end, a NS with hyperon seems to be a possibility. 
Same feature is evident in the gravitational mass evolution (Fig. 11) also. The 
full lines are for $40(23)M_{\odot}$ progenitors with $f_{heat}=1$. The 
other lines are for $f_{heat}>1$ and $23M_{\odot}$ progenitor. 
Therefore, we can conclude that $\Lambda$ hyperons alone are insufficient to 
support metastability, accretion of matter on PNS is also important for BH 
formation.  

\section{Summary}
We studied the effect of hadron-hyperon phase transition in core-collapse supernova using general relativistic hydrodynamic simulation GR1D \cite{Ott}. By following the dynamical collapse of a new-born proto-neutron star from the gravitational collapse of several progenitor stars adopting Shen hyperonic EoS table
\cite{shenY}, we noticed that $\Lambda$ hyperons appear just after bounce. It 
appears off center at first due to high temperature and prevails at the center of the PNS 
just before the BH formation, when the density becomes quite 
high. The $40M_{\odot}$ progenitor has been studied by several other authors \cite{Ott, ishi, oertel} due to its relatively large iron core.
We compare our results for the $40M_{\odot}$ progenitor with Sumiyoshi et al 
\cite{ishi}, who found $t_{BH}=0.682$s with their EoS containing $\Lambda$, 
$\Sigma$ and $\Xi$ hyperons. Peres et al.\cite{oertel} however found 
$t_{BH}=0.2745$s for LS220 + $\Lambda$ EoS. For Shen np$\Lambda$ EoS we 
observed BH formation at 0.565s.

$\Lambda$ hyperons triggers the black-hole formation, but fails to generate 
the second shock as the EoS is softened too much
with their appearance. The emergence of $\Lambda$ hyperon in the collapse 
produces an 
intense but short neutrino burst, which terminates at the BH formation. 
However, no second neutrino burst is observed as in quark-hadron phase 
transition \cite{sagert11}. The quark EoS is stiffer compared to hyperon EoS, which might be a possible reason for no-show of a second neutrino burst in the latter case. Also, accurate determination of neutrino fluxes and energy demands the use of better transport calculations than the schemes adopted in GR1D.

We reported the long-time evolution of the $23M_{\odot}$ progenitor in order to explain the observations following the advent of supernova SN1987A. We noticed in models with hyperonic EOS, the
maximum gravitational mass of the PNS exceeds that of the NS. The PNS cools via neutrinos and once sufficient thermal support is lost, no stable configuration exists and the PNS might collapse to a BH. Since one-dimensional supernova simulations is unlikely to produce a successful
explosion, we artificially adjusted the amount of neutrino heating via the 
scale factor $f_{heat}$ to achieve explosions. However, no BH formation was
noticed even after 20s(shown upto 11sec in Figs. 8-11). We find that the 
supernova ends up as a NS for $23M_{\odot}$ progenitors. 
For the metastability of the PNS and the delayed BH formation it is noted 
that hyperons alone could not do this job.

Also, the maximum mass in Shen np${\Lambda}$ EoS is not at par with the latest 
benchmark observations of neutron star masses \cite{demo,anto}. More hyperon 
physics at high densities required at this point. Still, existence of high mass neutron stars with hyperons is possible in quark-meson coupling model, SU(3) nonlinear sigma model, extended RMF model etc \cite{sch1,schaf1,schaf2}. We are 
working towards an EoS table with hyperons for supernova simulation with 
density-dependent couplings \cite{ddrh, typel}.

There are possibilities for other strange degrees of freedom in the form of kaon condensates to appear in the highly dense matter. We have seen such a phase transition can support a maximum mass \cite{meta}, which is well above 
$2M_{\odot}$ \cite{demo,anto}. It would be intriguing to investigate if a hadron-antikaon condensed matter can generate  the second shock and lead to
supernova explosion. A successful shock revival would have observational 
consequence in the form of neutrino signatures. Until now, only one supernova, SN1987A, has been detected by its neutrinos. Post SN1987A, more advanced neutrino facilities, such as ice-cube and super-Kamiokande are expected to detect the neutrino signals more efficiently and frequently.

\section{Acknowledgement}
It is a pleasure to thank 
Dr Evan O'connor and Dr. Christian Ott for immense help with
their code. Also, thanks to Prof. Debades Bandyopadhyay for 
insightful discussions and his meticulous feedback on the
manuscript.

\newpage 
\vspace{-2cm}

{\centerline{
\epsfxsize=12cm
\epsfysize=14cm
\epsffile{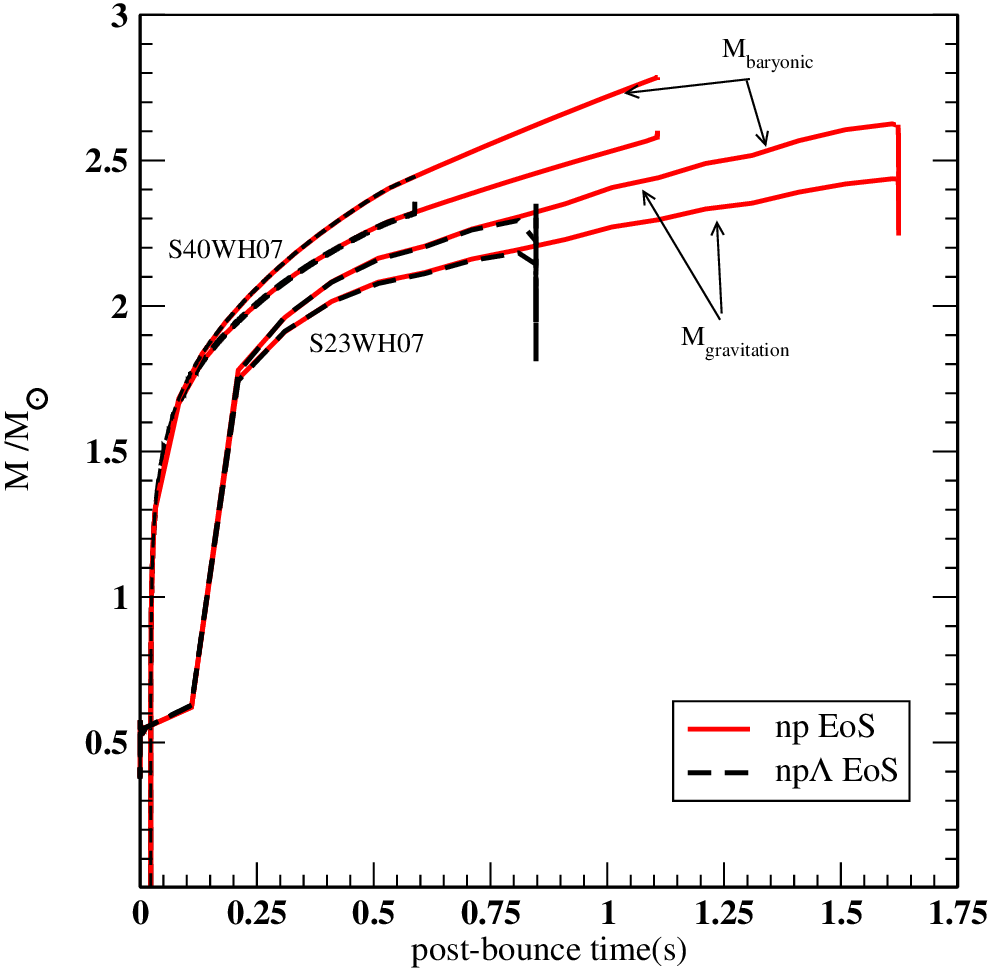}
}}

\vspace{4.0cm}

\noindent{\small{
FIG. 1. Temporal evolution of baryonic and gravitational mass for Shen np and np$\Lambda$ EoS.
}}

\newpage 
\vspace{-2cm}

{\centerline{
\epsfxsize=12cm
\epsfysize=14cm
\epsffile{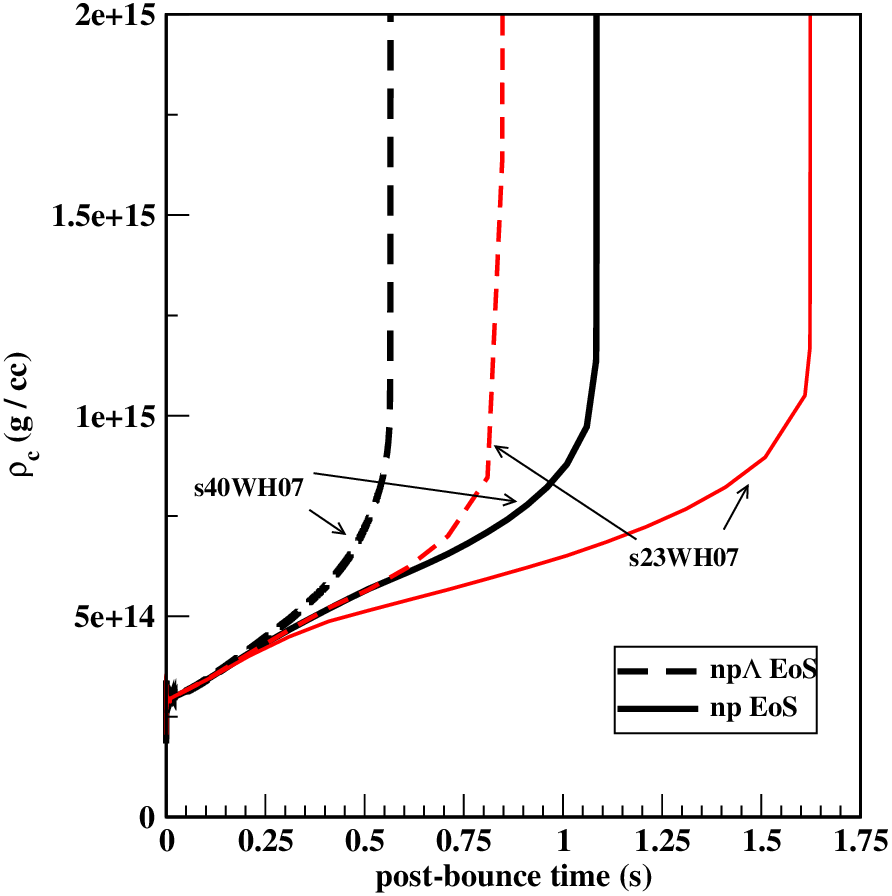}
}}

\vspace{4.0cm}

\noindent{\small{
FIG. 2. Temporal evolution of central density  for Shen np and np$\Lambda$ EoS.

\newpage 
\vspace{-2cm}

{\centerline{
\epsfxsize=12cm
\epsfysize=14cm
\epsffile{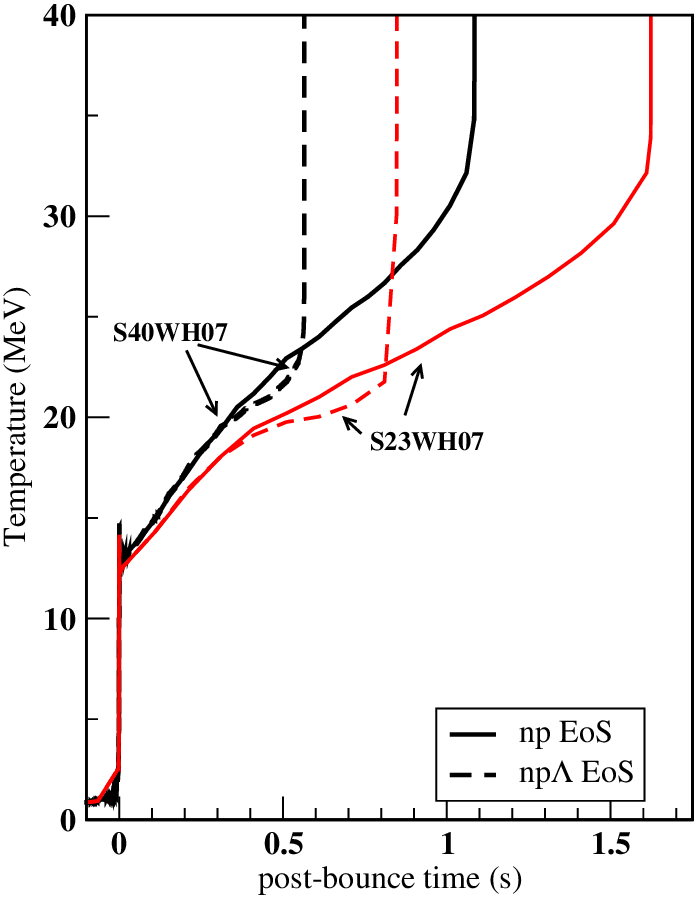}
}}

\vspace{4.0cm}

\noindent{\small{
FIG. 3. Temporal Evolution of temperature for Shen np and np$\Lambda$ EoS.
\newpage 
\vspace{-2cm}

{\centerline{
\epsfxsize=14cm
\epsfysize=12cm
\epsffile{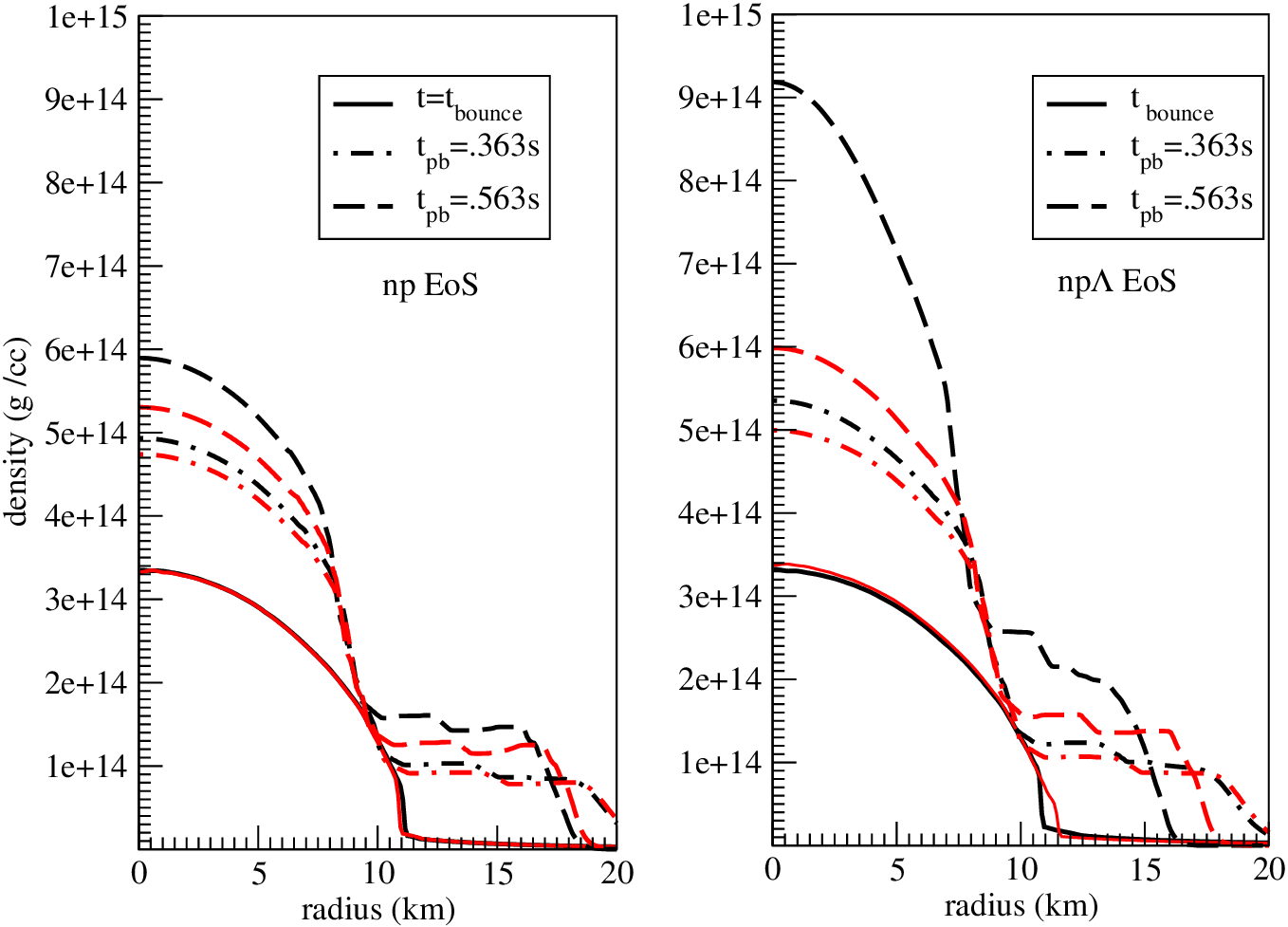}
}}

\vspace{4.0cm}

\noindent{\small{
FIG. 4. Density profile of the 40$M_{\odot}$(black lines) and 23$M_{\odot}$(colour online) 
progenitors with np (left panel)and np$\Lambda$ (right panel)EoS at 
$t=t_{bounce}$, and post-bounce time 0.363s and 0.563s.

\newpage 
\vspace{-2cm}

{\centerline{
\epsfxsize=14cm
\epsfysize=12cm
\epsffile{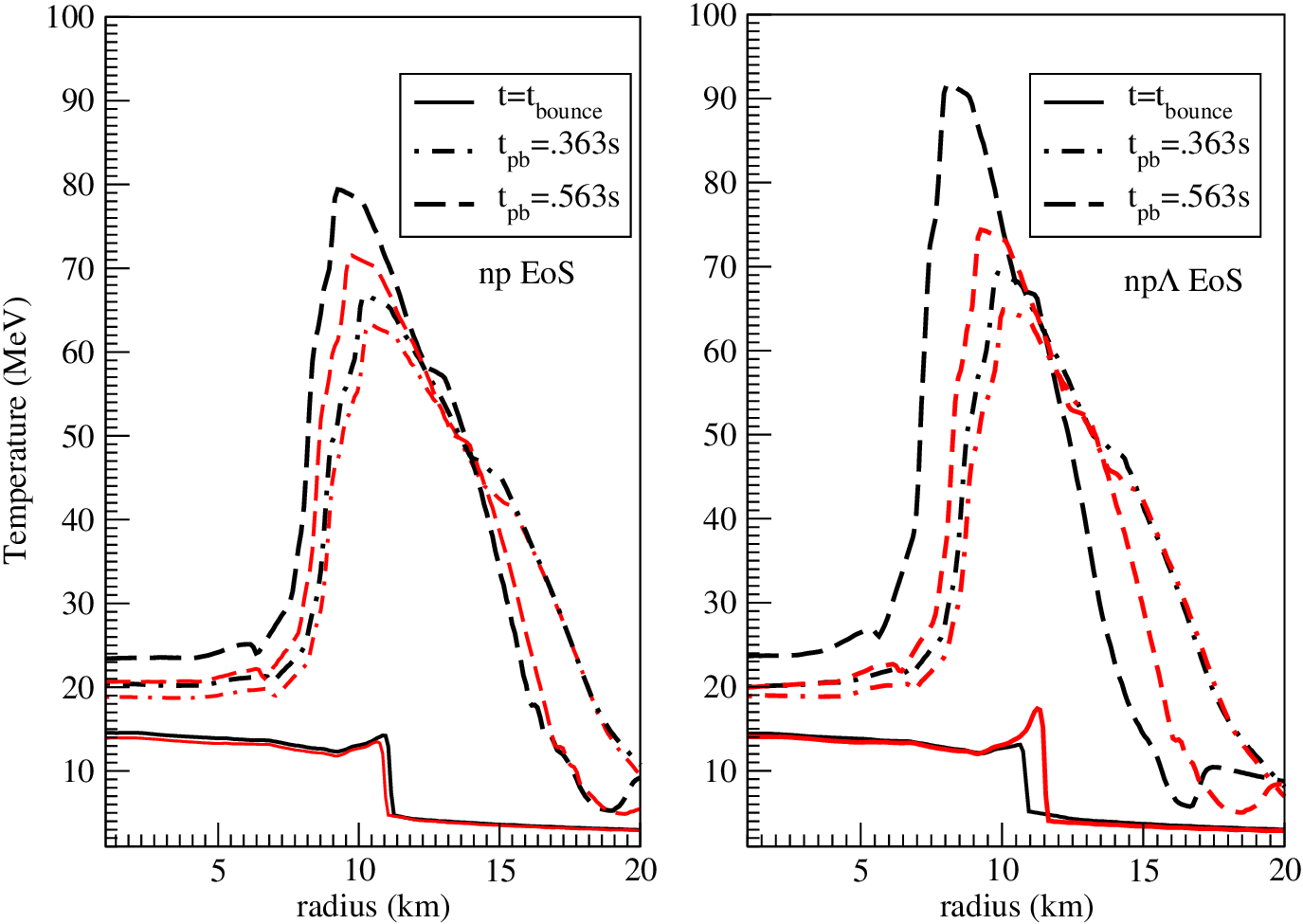}
}}

\vspace{4.0cm}

\noindent{\small{
FIG. 5. Temperature profile of the 40$M_{\odot}$(black lines) and 23$M_{\odot}$(colour online) progenitors with np (left panel)and np$\Lambda$ (right panel) 
EoS at $t=t_{bounce}$, and post-bounce time 0.363s and 0.563s.
}}

\newpage 
\vspace{-2cm}

{\centerline{
\epsfxsize=12cm
\epsfysize=14cm
\epsffile{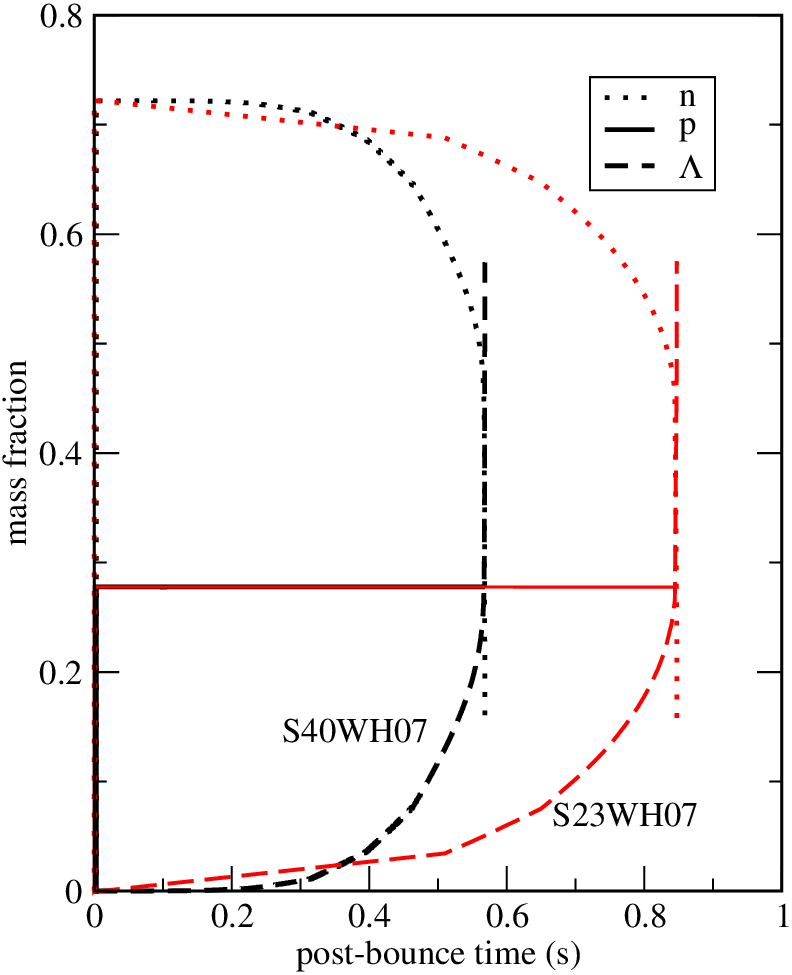}
}}

\vspace{4.0cm}

\noindent{\small{
FIG. 6. Mass fractions of the constituents for the 40$M_{\odot}$(black lines) and 23$M_{\odot}$(colour online) progenitors with Shen np$\Lambda$ EoS.
}}
\newpage 
\vspace{-2cm}

{\centerline{
\epsfxsize=14cm
\epsfysize=12cm
\epsffile{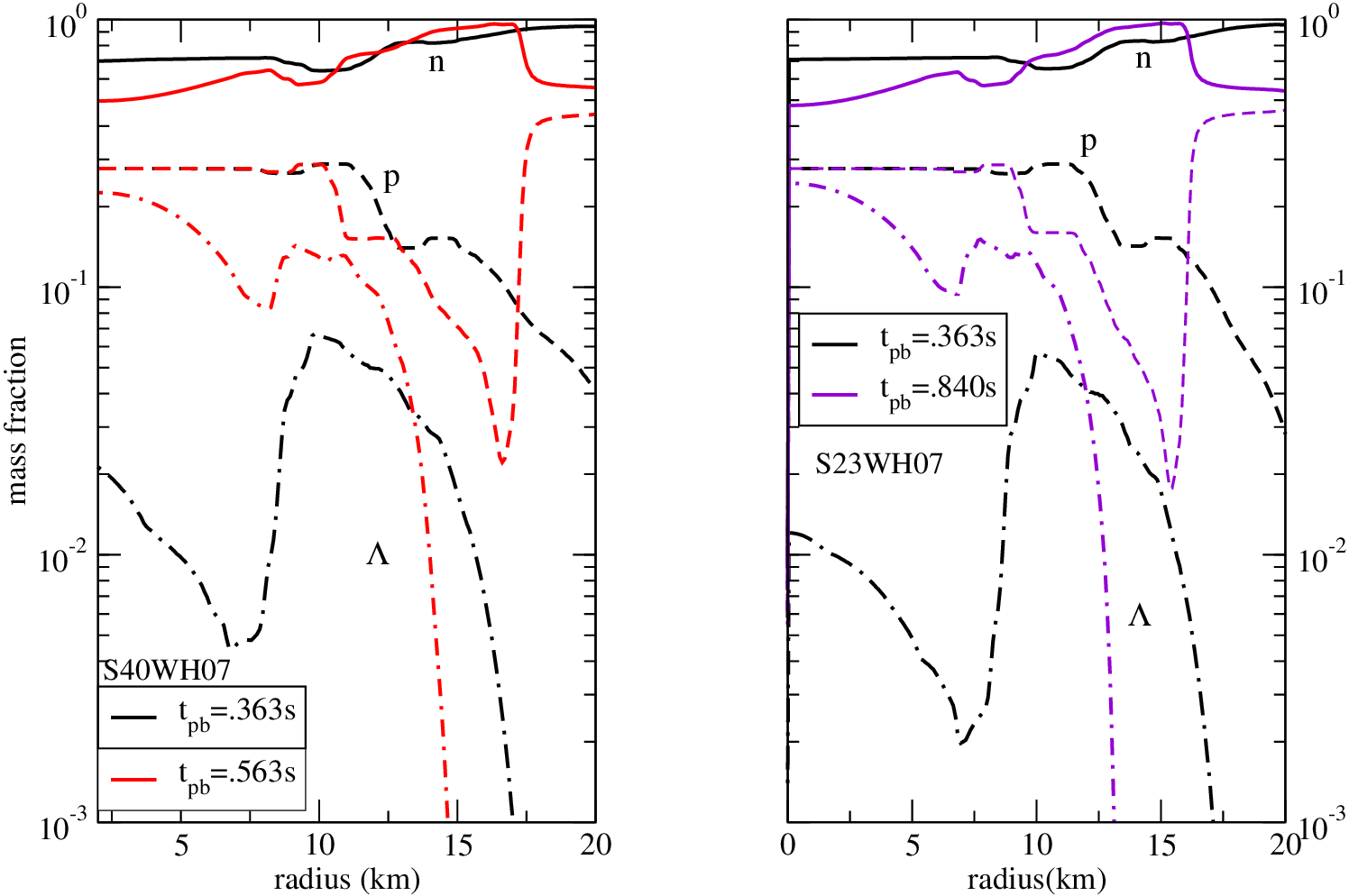}
}}

\vspace{4.0cm}

\noindent{\small{
FIG. 7. Snapshots of mass fractions of the constituents vs radius at 
t=0.363s(black lines) and 0.563s(colour online) post-bounce for the 
40$M_{\odot}$(left panel) and t=0.363s(black lines) and 0.84s(colour online) 
post-bounce for the 23$M_{\odot}$(right panel) progenitors with 
Shen np$\Lambda$ EoS.
}}
\newpage 
\vspace{-2cm}

{\centerline{
\epsfxsize=12cm
\epsfysize=14cm
\epsffile{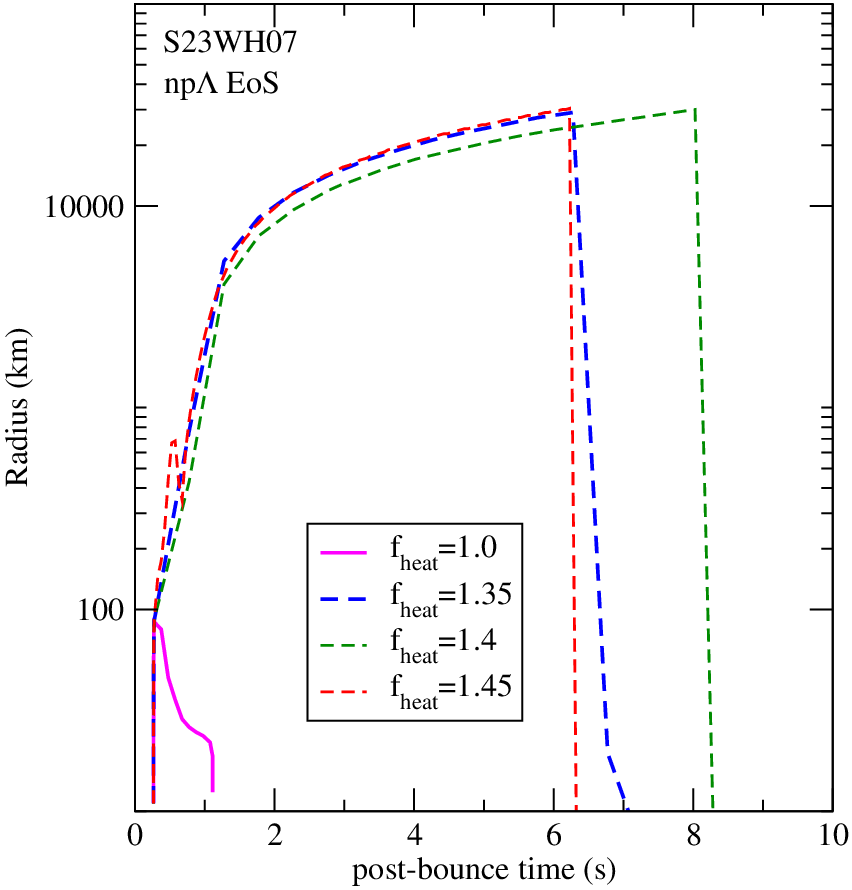}
}}

\vspace{4.0cm}

\noindent{\small{
FIG. 8. Evolution of shock radii for the $23M_{\odot}$ progenitor models.
}}

\newpage 
\vspace{-2cm}

{\centerline{
\epsfxsize=12cm
\epsfysize=14cm
\epsffile{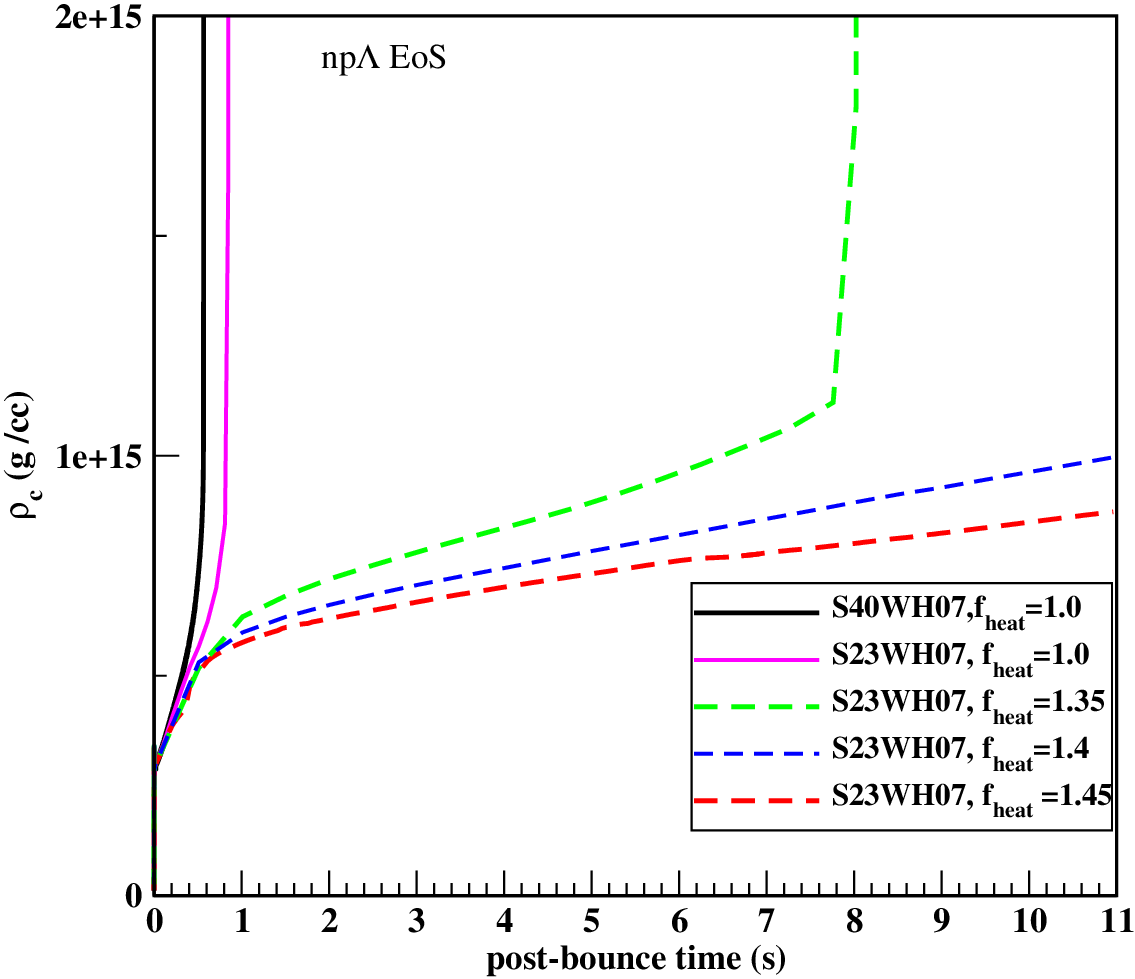}
}}

\vspace{4.0cm}

\noindent{\small{
FIG. 9. Long time evolution of  central density for the $23M_{\odot}$  progenitor model with np$\Lambda$ EoS.

\newpage 
\vspace{-2cm}

{\centerline{
\epsfxsize=12cm
\epsfysize=14cm
\epsffile{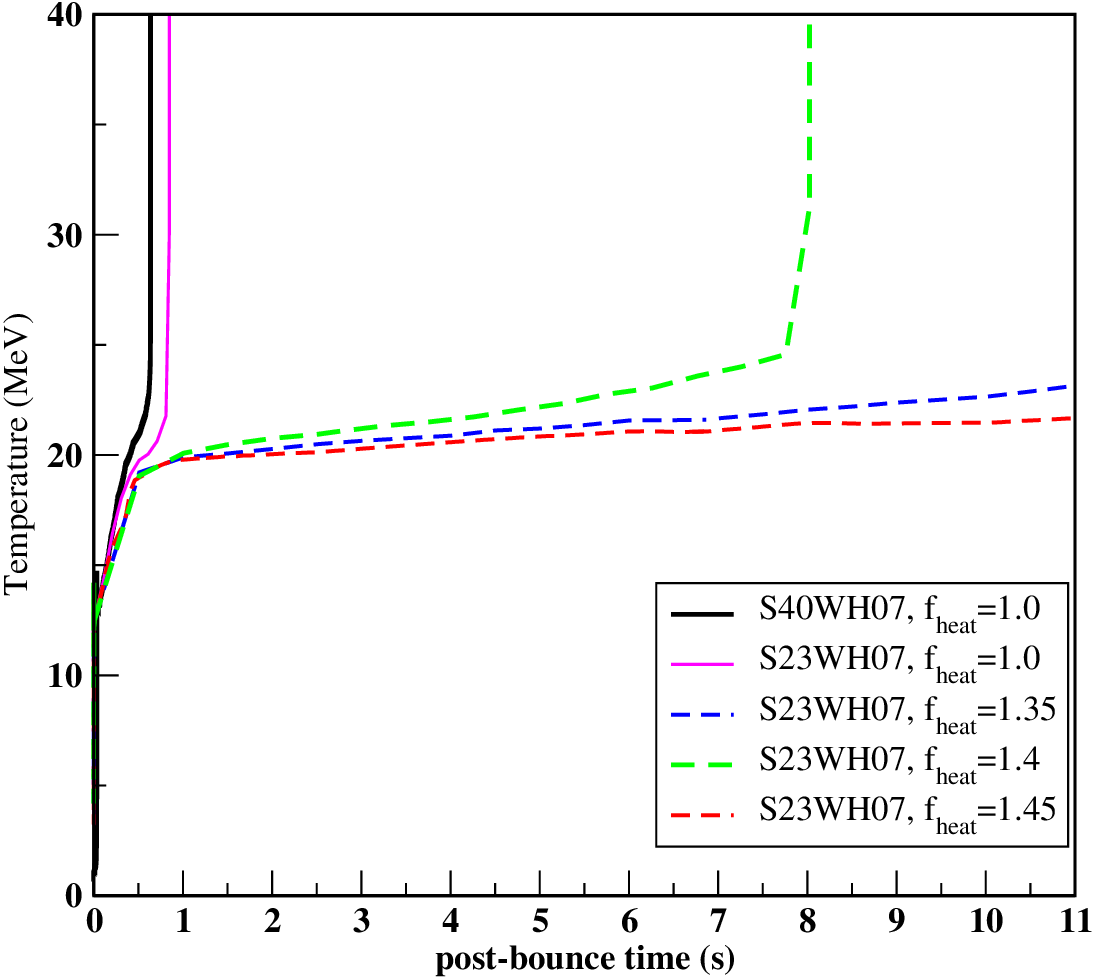}
}}

\vspace{4.0cm}

\noindent{\small{
FIG. 10. Long time evolution of  temperature for the $23M_{\odot}$  progenitor model with np$\Lambda$ EoS.
\newpage 
\vspace{-2cm}

{\centerline{
\epsfxsize=12cm
\epsfysize=14cm
\epsffile{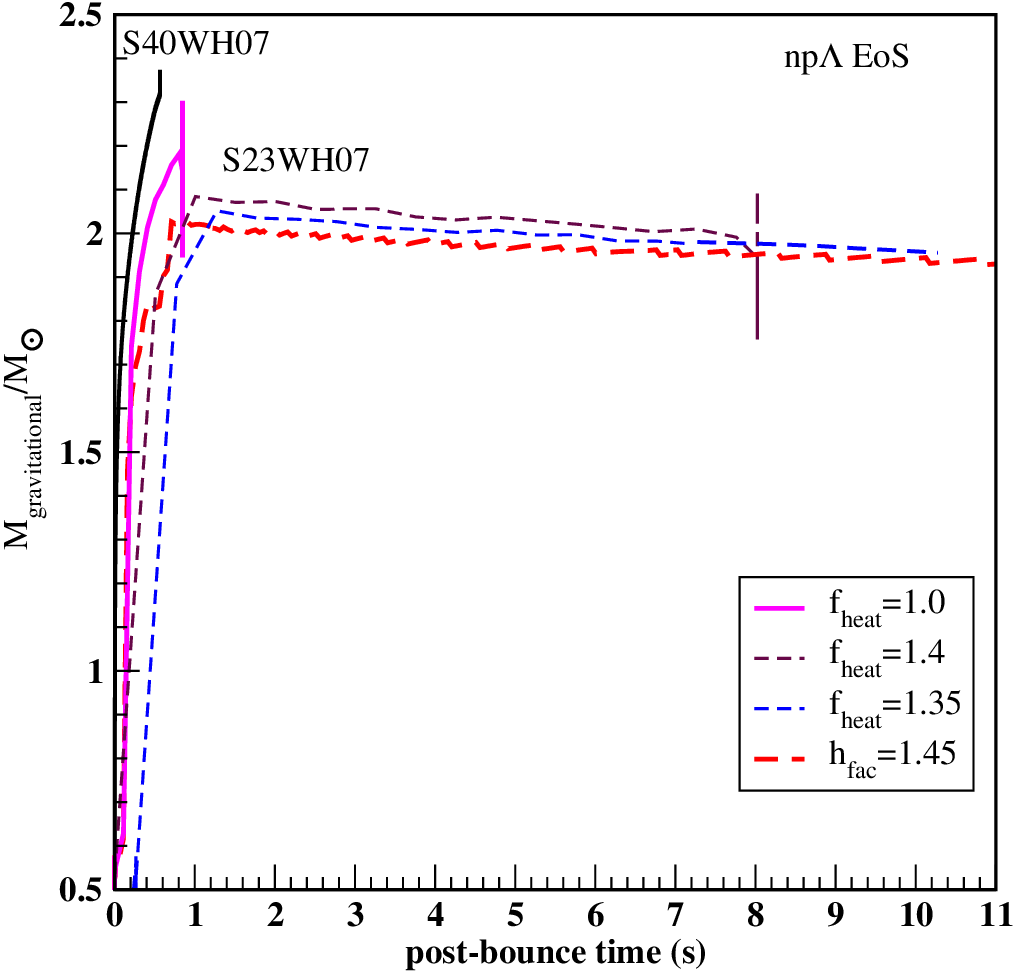}
}}

\vspace{4.0cm}

\noindent{\small{
FIG. 11. Long time evolution of  gravitational mass  for the $23M_{\odot}$  
progenitor model with np$\Lambda$ EoS. The dark full line is with $f_{heat}=1$ 
for $40M_{\odot}$, rest are for $23M_{\odot}$ progenitor.}}

\begin{thebibliography}{99}
\bibitem {Gle} N.K. Glendenning, Compact stars, (Springer, New York, 1997).
\bibitem{demo}P. B. Demorest, T. Pennucci,S. M. Ransom, M. S. E. Roberts \& J. W. T. Hessels, Nature {\bf 467} 1081 (2010).
\bibitem{anto} J. Antoniadis et. al., Science 26  {\bf 340} 6131 (2013). 
\bibitem{shenY}H. Shen,  H. Toki, K. Oyamatsu, K. Sumiyoshi, ApjS, {\bf 197} 20 (2011).
\bibitem{Lat10} J. M. Lattimer, M. Prakash, arXiv:1012.3208
\bibitem{mass}E. Massot, J. Margueron, and G. Chanfray, EPL {\bf 97} 39002 (2012).
\bibitem{oertel}B. Peres, M. Oertel, J. Novak, Phys. Rev. {\bf D 87}, 043006 (2013).
\bibitem{Lat}J. M. Lattimer and F. D. Swesty, Nucl. Phys. A {\bf 535} 331 (1991).
\bibitem{shen} H. Shen, H. Toki, K. Oyamatsu, K. Sumiyoshi, Nucl.Phys. {\bf A637} 435 (1998).
\bibitem{Sug} Y. Sugahara and H. Toki, Nucl. Phys. {\bf A579} 557 (1994).
\bibitem{hempel} M. Hempel and J. Schaffner-Bielich, Nucl. Phys. {\bf A 837}, 210-254 (2010).
\bibitem{furusawa} S. Furusawa, S. Yamada, K. Sumiyoshi, H. Suzuki, Astrophys.J. {\bf 738} 178 (2011).
\bibitem{horeq}G. Shen, C. J. Horowitz, S. Teige, Phys.Rev.{\bf C83}:035802,(2011).
\bibitem{ishi08}C. Ishizuka, A. Ohnishi, K. Tsubakihara, K. Sumiyoshi, S.Yamada, J. Phys. {\bf G35}  085201 (2008).
\bibitem{ishi}K. Sumiyoshi, C. Ishizuka, A. Ohnishi,  S. Yamada, H. Suzuki, APjL,{\bf 690}, L43 (2009).
\bibitem{naka11}K. Nakazato, S. Furusawa, K. Sumiyoshi, A. Ohnishi, S. Yamada, H. Suzuki, Astrophys.J.{\bf745} 197 (2012).
\bibitem {bethe} H. A. Bethe, Rev. Mod. Phys. {\bf 62} 801 (1990).
\bibitem{sagert11} T. Fischer, I. Sagert, G. Pagliara, M. Hempel, J. Schaffner-Bielich, T. Rauscher, F.-K. Thielemann, R. Käppeli, G. Martínez-Pinedo, M. Liebendörfer ApJ.Supp. {\bf 194}:39, (2011). 
\bibitem{meta}S. Banik and D. Bandyopadhyay, Phys. Rev. {\bf C63} 035802 (2001).
\bibitem{brown}G.E. Brown and H.A. Bethe, Astrophys. J. {\bf 423}, 659 (1994). 
\bibitem{ruff} C. Rhoades, Jr. and R. Ruffini. Phy. Rev. Lett, {\bf 32}, 324. (1974).
\bibitem{brown2} G.E. Brown, S.W. Bruenn and J.C. Wheeler, Comments Astrophys. {\bf 16}, 153 (1992).
\bibitem{prak_meta}M. Prakash, J.R. Cooke and J.M. Lattimer, Phy Rev {\bf D52} 661 (1995).
\bibitem{third} S. Banik and D. Bandyopadhyay, Phys. Rev. {\bf C64} 055805 (2001). 
\bibitem{schaf} J. Schaffner, I. Mishustin, Phys. Rev{\bf C 53}, 4 (1995).
\bibitem{Ott} E. O'Connor, C. D. Ott, ApJ {\bf 730} 70 (2011).
\bibitem{Keil}W. Keil, H.T. Janka, Astron. and Astrophys, {\bf 296}, 145 (1995).
\bibitem{Baum} T. W. Baumgarte, H.T. Janka, W. Keil, S. L. Shapiro, 
S.A. Teukolsky, Astrophys J. {\bf 468}, 823 (1996).
\bibitem{shen06}H. Shen,F. Yang, H. Toki, Prog. Theo. Phys., {\bf 115} 325 (2006).
\bibitem{Lieb} Matthias  Liebend\"orfer, Astrophys.J. {\bf 633} 1042 (2005)
\bibitem{janka}H.-Th. Janka, A \& A, {\bf 368}, 527 (2001)
\bibitem{woo}S. E. Woosley, A. Heger, Phys. Rept. {\bf 442}:269-283,(2007).
\bibitem{ddrh}S. Banik and D. Bandyopadhyay, Phys. Rev. {\bf C66} 065801 (2002).
\bibitem{ott_fail} Evan O'Connor, Christian D. Ott, arXiv:1103.2117v1 
\bibitem{JPCS} S. Banik,  J. of Phys: Conf. Series, {\bf426}, Number 1, 12004 (2013).
\bibitem{sch1}V. Dexheimer, S. Schramm, Astrophys.J.{\bf 683}:943-948,(2008).
\bibitem{schaf1}S. Weissenborn, D. Chatterjee, J. Schaffner-Bielich, Phys. Rev. {\bf C 85}, 065802 (2012).
\bibitem{schaf2}S. Weissenborn, D. Chatterjee, J. Schaffner-Bielich, Nucl. Phys. {A 881} 62 (2012).
\bibitem{typel}S. Typel, G. R\"opke, T. Kl\"ahn, D. Blaschke and H.H. Wolter, 
Phys. Rev.{\bf C 81} 015803,(2010).
\end{thebibliography}
\end{document}